\documentclass[aps,prl,reprint,superscriptaddress]{revtex4-1}

\usepackage{graphicx}
\usepackage[utf8x]{inputenc}
\usepackage{amsmath}
\usepackage[]{xcolor}
\DeclareMathOperator{\SU}{SU}
\DeclareMathOperator{\U}{U}

\begin{document}

\def\be{\begin{equation}}
\def\ee{\end{equation}}
\def\bea{\begin{eqnarray}}
\def\eea{\end{eqnarray}}

\def\reff#1{\ref{#1}}
\def\labell#1{\label{#1}}

\def\app#1{Appendix~\reff{#1}}
\def\apps#1#2{Appendices~\reff{#1}--\reff{#2}}
\def\eq#1{Eq.~(\reff{#1})}
\def\eqs#1#2{Eqs.~(\reff{#1})--(\reff{#2})}
\def\eqss#1#2#3{Eqs.~(\reff{#1}), (\reff{#2}) and (\reff{#3})}
\def\fig#1{Fig.~\reff{#1}}
\def\figs#1#2{Figs.~\reff{#1}--\reff{#2}}
\def\sec#1{Sec.~\reff{#1}}
\def\tab#1{Table~\reff{#1}}
\def\tabs#1#2{Tabs.~\reff{#1}--\reff{#2}}

\newcommand{\lsim}{ {\
\lower-1.2pt\vbox{\hbox{\rlap{$<$}\lower5pt\vbox{\hbox{$\sim$}}}}\ } }
\newcommand{\gsim}{ {\
\lower-1.2pt\vbox{\hbox{\rlap{$>$}\lower5pt\vbox{\hbox{$\sim$}}}}\ } }

\def\dof{\mathrm{dof}}
\def\er#1#2{\relax\ifmmode{}^{+#1}_{-#2}\else$^{+#1}_{-#2}$\fi}
\def\erparen#1#2{\relax\ifmmode{}(^{#1}_{#2})\else$(^{#1}_{#2})$\fi}
%

% New ~. In math mode gives space as wide as a zero, in text
% mode it's a tie, as usual
\def~{\ifmmode\phantom{0}\else\penalty10000\ \fi}
\def\half{{1\over2}}
\def\bra#1{\left\langle #1\right|}
\def\ket#1{\left| #1\right\rangle}
\def\vev#1{\left\langle #1\right\rangle}
\def\hc{\mathrm{h.c.}}
\def\su#1{{SU\left(#1\right)}}
\def\u#1{{U\left(#1\right)}}
\def\d#1#2{d\mskip 1.5mu^{#1}\mkern-1mu{#2}\,}
\def\D#1#2{{\d#1{#2} \over (2\pi)^{#1}}\,}
\def\dtilde#1{{\d3{#1} \over (2\pi)^3 \,2\omega_{#1}}\,}
\def\tr{\mathrm{tr}}
\def\Tr{\mathrm{Tr}}
\def\str{\mathrm{str}}
\def\det{\mathrm{det}}
\def\sdet{\mathrm{sdet}}
\def\fm{\mathrm{fm}}
\def\ev{\mathrm{e\kern-0.1em V}}
\def\kev{\mathrm{ke\kern-0.1em V}}
\def\mev{\mathrm{Me\kern-0.1em V}}
\def\gev{\mathrm{Ge\kern-0.1em V}}
\def\tev{\mathrm{Te\kern-0.1em V}}
\def\ps{\mathrm{ps}}
\def\re{\mathrm{Re}}
\def\im{\mathrm{Im}}
\let\Re=\re \let\Im=\im
\def\dotp#1#2{#1\mathord\cdot #2}
\def\n#1e#2n{{#1}\times 10^{#2}}

\def\ra{\rangle}
\def\la{\langle}
\def\l{\left}
\def\r{\right}
\def\ord#1{\mathcal{O}(#1)}
\def\nn{\nonumber}
\def\On{\mathcal{O}_n}

\def\cO{\mathcal{O}}
\def\cF{\mathcal{F}}
\def\cP{\mathcal{P}}
\def\cH{\mathcal{H}}
\def\cL{\mathcal{L}}
\def\cM{\mathcal{M}}
\def\cD{\mathcal{D}}
\def\cE{\mathcal{E}}
\def\cR{\mathcal{R}}
\def\cpv{
\math{\mathrm{CP}\hspace{-0.6cm}\slash{}\hspace{0.4cm}}}
\def\ods2{\mathcal{O}_{\Delta S=2}}
\def\zds2{Z_{\Delta S=2}}
\def\dslash{\partial\!\!\!\slash}

\def\msbar{{\overline{\mathrm{MS}}}}
\def\NDR{\mathrm{NDR}}
\def\RI{\mathrm{RI}}
\def\RGI{\mathrm{RGI}}
\def\BSM{\mathrm{BSM}}
\def\LO{\mathrm{LO}}
\def\NLO{\mathrm{NLO}}
\def\AWI{\mathrm{AWI}}
\def\lqcd{\Lambda_\mathrm{QCD}}
\def\lat{\mathrm{lat}}
\def\qed{\mathrm{QED}}
\def\qcd{\mathrm{QCD}}
\def\exp{\mathrm{exp}}

\makeatletter
\def\slash#1{{\mathpalette\c@ncel{#1}}} % TeXbook, bottom of p360
% put back plain TeX's \big etc macros
\def\big#1{{\hbox{$\left#1\vbox to1.012\ht\strutbox{}\right.\n@space$}}}
\def\Big#1{{\hbox{$\left#1\vbox to1.369\ht\strutbox{}\right.\n@space$}}}
\def\bigg#1{{\hbox{$\left#1\vbox to1.726\ht\strutbox{}\right.\n@space$}}}
\def\Bigg#1{{\hbox{$\left#1\vbox
to2.083\ht\strutbox{}\right.\n@space$}}}
\makeatother

\newcommand{\lonebar}{\ln\frac{\Lambda_1^2}{M_\pi^2}}
\newcommand{\ltwobar}{\ln\frac{\Lambda_2^2}{M_\pi^2}}
\newcommand{\lthreebar}{\ln\frac{\Lambda_3^2}{M_\pi^2}}
\newcommand{\lfourbar}{\ln\frac{\Lambda_4^2}{M_\pi^2}}
\newcommand{\lsixbar}{\ln\frac{\Lambda_6^2}{M_\pi^2}}
\newcommand{\lMbar}{\ln\frac{\Omega_M^2}{M_\pi^2}}
\newcommand{\lFbar}{\ln\frac{\Omega_F^2}{M_\pi^2}}
\newcommand{\lSbar}{\ln\frac{\Omega_S^2}{M_\pi^2}}

\newcommand{\phys}{\mathrm{ph}}
\newcommand{\awi}{\mathrm{AWI}}
\newcommand{\vwi}{\mathrm{VWI}}
\newcommand{\pcac}{\mathrm{PCAC}}
\newcommand{\imp}{\mathrm{imp}}
\newcommand{\bare}{\mathrm{bare}}

\newcommand{\cverrlist}[5]{&=& #1(#2)_\mathrm{stat}
                         \left(_{#3}^{#4}\right)_\mathrm{sys}
                         \,#5}

\newcommand{\errlist}[3]{#1(#2)%_\mathrm{stat}
                         (#3)%_\mathrm{sys}
}

\def\T4{\mathbb{T}^4}
\def\prodmu{\prod_{\mu=0}^3}
\def\BZT4{\mathrm{BZ}(\mathbb{T}^4)}

%\preprint{}

\title{Isospin Splittings in the Light-Baryon Octet from Lattice QCD and QED}


%\author{}
%\thanks{}
%\affiliation{}

%% affiliations
\newcommand{\budapesta}{\affiliation{Institute for Theoretical Physics, E\"otv\"os
    University, P\'azm\'any Peter s\'etany 1/A, H-1117 Budapest, Hungary}}
\newcommand{\budapestb}{\affiliation{MTA-ELTE Lend\"ulet Lattice Gauge Theory Research Group, Budapest, Hungary}}
\newcommand{\juelich}{\affiliation{IAS/JSC, Forschungszentrum J\"ulich, D-52425
    J\"ulich, Germany}}
\newcommand{\marseille}{\affiliation{Aix-Marseille Université, CNRS, CPT, UMR
    7332, 13288 Marseille, France}}
\newcommand{\toulon}{\affiliation{Université de Toulon, CNRS, CPT, UMR 7332,
    83957 La Garde, France}}
\newcommand{\wuppertal}{\affiliation{Department of Physics, Wuppertal
    University, Gaussstrasse 20, D-42119 Wuppertal, Germany}}

%\author{Szabolcs Borsanyi}
\author{Sz.~Borsanyi}
\wuppertal

%\author{Stephan D\"urr}
\author{S.~D\"urr}
\wuppertal
\juelich

%\author{Zoltan Fodor}
\author{Z.~Fodor}
\wuppertal
\juelich
\budapesta

%\author{Julien Frison} 
\author{J.~Frison} \altaffiliation{Present address: School of Physics
  and Astronomy, University of Edinburgh, Edinburgh EH9 3JZ, UK}
\marseille
\toulon

%\author{Christian Hoelbling}
\author{C.~Hoelbling}
\wuppertal

%\author{Sandor Katz}
\author{S.D.~Katz}
\budapesta
\budapestb

%\author{Stefan Krieg}
\author{S.~Krieg}
\wuppertal
\juelich

%\author{Thorsten Kurth}
\author{Th.~Kurth}
\wuppertal

%\author{Laurent Lellouch}
\author{L.~Lellouch}
\marseille
\toulon

%\author{Thomas Lippert}
\author{Th.~Lippert}
\juelich

%\author{Antonin Portelli}
\author{A.~Portelli}
\marseille
\toulon
\affiliation{School of Physics and Astronomy, University of Southampton, 
Southampton SO17 1BJ, United Kingdom}

%\author{Alberto Ramos}
\author{A.~Ramos}
\altaffiliation{Present address: NIC, DESY Platanenallee 6, D-15738 Zeuthen, 
Germany}
\marseille
\toulon

%\author{Alfonso Sastre}
\author{A.~Sastre}
\marseille
\toulon

%\author{Kalman Szabo}
\author{K.~Szabo}
\wuppertal

\collaboration{Budapest-Marseille-Wuppertal Collaboration}

\date{\today}

\begin{abstract}
While electromagnetic and up-down quark mass difference effects on octet baryon
masses are very small, they have important consequences. The stability of the
hydrogen atom against beta decay is a prominent example. Here we include these
effects by adding them to valence quarks in a lattice QCD calculation based on
$N_f{=}2{+}1$ simulations with 5 lattice spacings down to 0.054~fm, lattice
sizes up to 6~fm and average up-down quark masses all the way down to their
physical value. This allows us to gain control over all systematic errors,
except for the one associated with neglecting electromagnetism in the sea. We
compute the octet baryon isomultiplet mass splittings, as well as the
individual contributions from electromagnetism and the up-down quark mass
difference. Our results for the total splittings are in good agreement with
experiment.

\end{abstract}

%\keywords{}

\maketitle

The existence and stability of atoms and ordinary matter rely heavily on the
fact that neutrons are slightly more massive than protons. The difference in
the mass of these two particles has been measured very precisely and is only
0.14\% of their average mass \cite{Beringer:1900zz}. Although it has yet to be
shown from first principles, we believe that this tiny difference results from
the competition between electromagnetic (EM) effects proportional to the fine
structure constant $\alpha\equiv e^2/(4\pi)$ and mass isospin breaking
effects proportional to the mass difference of up and down quarks $\delta
m\equiv m_u-m_d$. Here, we study this issue in the light-baryon octet. In
particular, we compute mass splittings in the nucleon ($N$), $\Sigma$, and $\Xi$
isospin multiplets using lattice QCD, to which we add QED in the valence quark
sector. Although one would also have to account for QED contributions from sea
quarks to have a complete calculation, these effects are suppressed, as
discussed below. Moreover, the approach taken here allows us to use a very rich
set of QCD gauge configurations that we have already generated
\cite{Durr:2008rw,Durr:2010vn,Durr:2010aw}. Eliminating the uncertainty
associated with neglecting QED sea-quark contributions would require performing
completely new simulations, implementing reweighting techniques
\cite{Ishikawa:2012ix,Aoki:2012st}, or using EM current insertion methods
\cite{deDivitiis:2013xla} and including quark-disconnected contributions. Such
a computation is beyond the scope of the present work.

Because mass and EM isospin symmetry breaking corrections are small and of
comparable size, it is legitimate to expand the standard model in powers of
$\delta m$ and $\alpha$, assuming $O(\delta m)\sim O(\alpha)$~\footnote{Here
and below it is assumed that in $O(\delta m^n)$, $\delta m$ is normalized by a
typical QCD mass scale.}. This expansion is expected to converge very rapidly,
with each subsequent order contributing $\sim 1\%$ of the previous one. Given
the size of other uncertainties in our calculation, we can safely work at LO in
this expansion, i.e. at $O(\delta m,\alpha)$.

\medskip
\noindent{\em The physical point.--} In the absence of weak interactions and
for energies smaller than the charm-anticharm threshold, the standard model of
quarks has five parameters that must be fixed by comparison to experiment. Here
we trade these parameters for observables which are particularly sensitive to
them: 1) the lattice spacing, $a$, for the mass of the decuplet baryon
$\Omega^-$--alternatively the isospin averaged $\Xi$ mass--as in
\cite{Durr:2008zz}, 2) the average $u$-$d$ mass, $m_{ud}$, for $M_{\pi^+}^2$,
3) the strange mass, $m_s$, for $M_{K^\chi}^2\equiv
(M_{K^+}^2+M_{K^0}^2-M_{\pi^+}^2)/2$, 4) $\delta m$ for the mass-squared
difference $\Delta M_K^2\equiv M_{K^+}^2-M_{K^0}^2$, and 5) bare $\alpha$ for
its renormalized value because it does not renormalize in our quenched QED
calculation. The {\em physical point} is then reached by tuning these
observables to their physical values given in \cite{Beringer:1900zz}, while
taking the continuum $a\to 0$ and infinite-volume $L\to\infty$ limits.

\medskip
\noindent
{\em Separating EM and $\delta m$ contributions.--} In addition to computing
the total splittings, it is interesting to separate them into a contribution
coming from $\delta m$ and one coming from QED. We define the EM
contribution by setting $\delta m=0$ via $\Delta M^2=M_{\bar uu}^2-M_{\bar
dd}^2=0$, with all other parameters tuned to their physical values. Here,
$M_{\bar qq}$ is the mass of a neutral meson $\bar qq$, $q=u,d$, whose
propagator includes only quark-connected diagrams. Using the $\chi$PT results
of \cite{Bijnens:2006mk}, it is straightforward to show that the difference of
these mesons' squared masses is $\Delta M^2=2 B_2\delta m + O(\alpha
m_{ud},\delta m m_{ud},\alpha\delta m,\alpha^2)$, where $B_2$ is the $N_f{=}2$
quark condensate parameter. Close to the physical point, $O(m_{ud})$ can be
counted like $O(\delta m)$. Thus, our definition of the EM contribution
differs from any other valid one by corrections of the size of NLO isospin
breaking terms, which are one order higher than the one to which we work here.
To obtain the contribution from $\delta m$, we set $\alpha=0$ and all other
parameters to their physical values. In particular, the physical value of
$\Delta M^2$ is obtained from the analysis of $\Delta M_K^2$ briefly described
below and by computing the value of $\Delta M^2$ corresponding to the physical
$\Delta M_K^2$. This analysis and its implications for Dashen's theorem
\cite{Dashen:1969eg} and $m_{u,d}$, which are very interesting in their own
right, will be discussed elsewhere \cite{bmwdashen}.

\medskip
\noindent
{\em Simulation details and parameters.--} We start from our 47, isospin
symmetric, $N_f{=}2{+}1$, QCD ensembles, obtained from simulations with pion
masses down to 120~MeV, lattice sizes up to 6~fm, five lattice spacings down to
0.054~fm, and more than $1000$ trajectories each \cite{Durr:2010aw}. To
these ensembles, we add QED gauge degrees of freedom in a way which has now
become standard \cite{Duncan:1996xy,Blum:2007cy,Portelli:2010yn}. For each QCD
gauge configuration, we generate an EM field $A_\mu(x)$ defined on the
links, using the noncompact EM action (in Coulomb gauge) and the methods
detailed in \cite{Blum:2007cy}. The gauge potential is then exponentiated as
$U_\mu^\qed(x) = \exp\left[iqe A_\mu(x)\right]$. Unlike the QCD links, our QED
links undergo no smearing before being coupled to quarks. Similarly, we have
not added a clover improvement term for the $\U(1)$ field. The $\U(1)$ fields
are then multiplied with the $\SU(3)$ gauge variable on each link and inserted
into the Wilson-Dirac operator associated with the quark of charge $q$ before
inversion. The resulting quark propagators are combined into meson and baryon
two-point functions. The extended sources and sinks used are the same as in
\cite{Durr:2010vn,Durr:2010aw}.

\begin{figure}[t]
\includegraphics[width=\columnwidth]{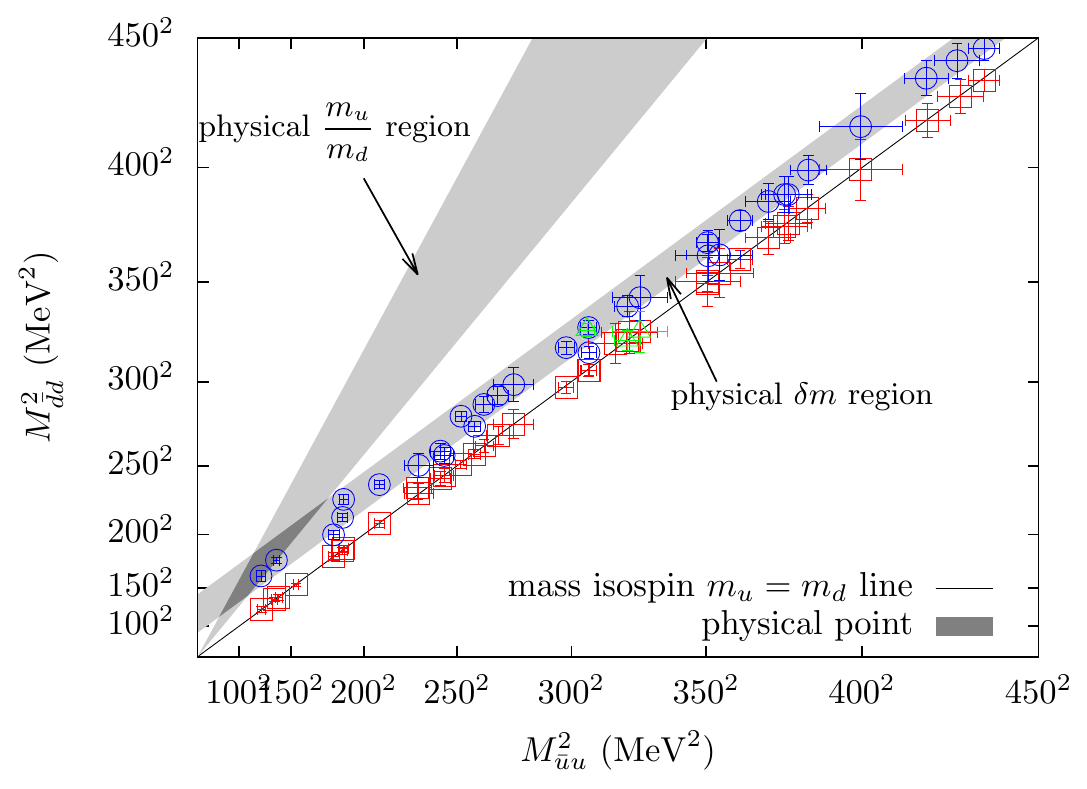}%
\caption{\label{fig:mu-md-landscape}\sl 
  Valence data sets plotted in the $M_{\bar dd}^2$ vs $M_{\bar uu}^2$ plane. The
  red squares (set 1) lie along the mass isospin line $M_{\bar uu}^2{=}M_{\bar
  dd}^2$, and the blue circles (set 2) are scattered around an estimate of the
  $\Delta M^{2,\phys}$ region, obtained from the results of
  \cite{Colangelo:2010et,Beringer:1900zz}. The green triangles (set 3) are
  points in which $\alpha$ is varied away from its physical value. For clarity,
  points with $M_\pi>450~\mev$ are not shown.}
\end{figure}

For most of our $\SU(3)$ ensembles, we have generated two valence data sets,
which include QED with the physical value of $\alpha$. In the first set (set 1),
the bare, valence $u$, $d$, and $s$ quark masses are individually tuned so that
their PCAC values approximately reproduce the corresponding ensemble's light
and strange sea-quark PCAC masses. Thus, we subtract the $\alpha/a$ divergences
in the valence bare quark masses, which come from the EM self-energy, as
described in \cite{Portelli:2010yn}. In the second set (set 2), we choose $m_d$
to be slightly more massive than in the first set so that $\Delta M^2$ scatters
around its physical value from ensemble to ensemble. We have one additional
valence data set (set 3) in which $\alpha$ is varied. The latter includes a
point with $\alpha\sim 2\alpha^\phys$ and $\Delta M^2\simeq \Delta
M^{2,\phys}$, a second with $\alpha\sim \alpha^\phys/4$ and a similar $\Delta
M^2$, and a third with $\alpha\sim 0$ and $\Delta M^2\simeq 0$. The superscript
$\phys$ indicates that we are referring to the physical value of a
quantity. We have 74 valence points in total, which are shown in the $M_{\bar
uu}^2$-$M_{\bar dd}^2$ plane in \fig{fig:mu-md-landscape}. This rich collection
of data allows us to gain full control over the dependence of the splittings on
all of the relevant parameters.

\medskip
\noindent {\em Analysis of meson and baryon correlators.--} The time dependence
of the $\pi^+$, $K^+$, $K^0$, and the $\Omega^-$ or $\Xi$ two-point functions is
fitted, in the asymptotic regime, to a hyperbolic cosine and an exponential,
respectively. For the isospin multiplets whose splittings we wish to determine,
we perform a simultaneous, correlated fit to the two-point functions of the two
members of the multiplet in which we replace the individual hadron mass
parameters by their average and their difference. The time ranges for
correlator fits are determined after a systematic study of the goodness of fit
as a function of initial and final fitting times. The choices made here are
very similar to those of \cite{Durr:2010vn,Durr:2010aw}.

\medskip
\noindent {\em Interpolating to the physical point and determining the
individual EM and $\delta m$ contributions.--} Having determined the isospin
splittings and relevant hadron masses in lattice units for each of our QCD plus
QED data sets, we have to convert them to physical units and extrapolate them to
the continuum and infinite-volume limits. We also must interpolate the
splittings to the physical mass point, as well as to the mass and EM isospin
limits.

We determine the five lattice spacings simultaneously from a combined fit of
the data with $\Delta M^2\simeq 0$, for the isospin symmetric observable
$aM_{\Omega^-}$ or, alternatively, $aM_\Xi$, using the techniques of
\cite{Durr:2008zz,Durr:2010aw,Durr:2010vn}. The isospin mass splitting $\Delta
M_X$ of a hadron $X$ is naturally described by the LO isospin expansion
\be
\label{eq:DMXisoexp}
\Delta M_X = A_X \alpha + B_X\Delta M^2
\ ,
\ee
where $\Delta M^2$ substitutes for $\delta m$. The coefficients $A_X$ and $B_X$
still depend on the isospin symmetric parameters of the theory, e.g.\ $m_{ud}$
or $m_s$. We find that their dependences on these parameters are well described
by a linear expansion in $M_{\pi^+}^2$ and in $M_{K^\chi}^2$ for the range of
masses retained below.

\begin{figure}[t]
\includegraphics[width=\columnwidth]{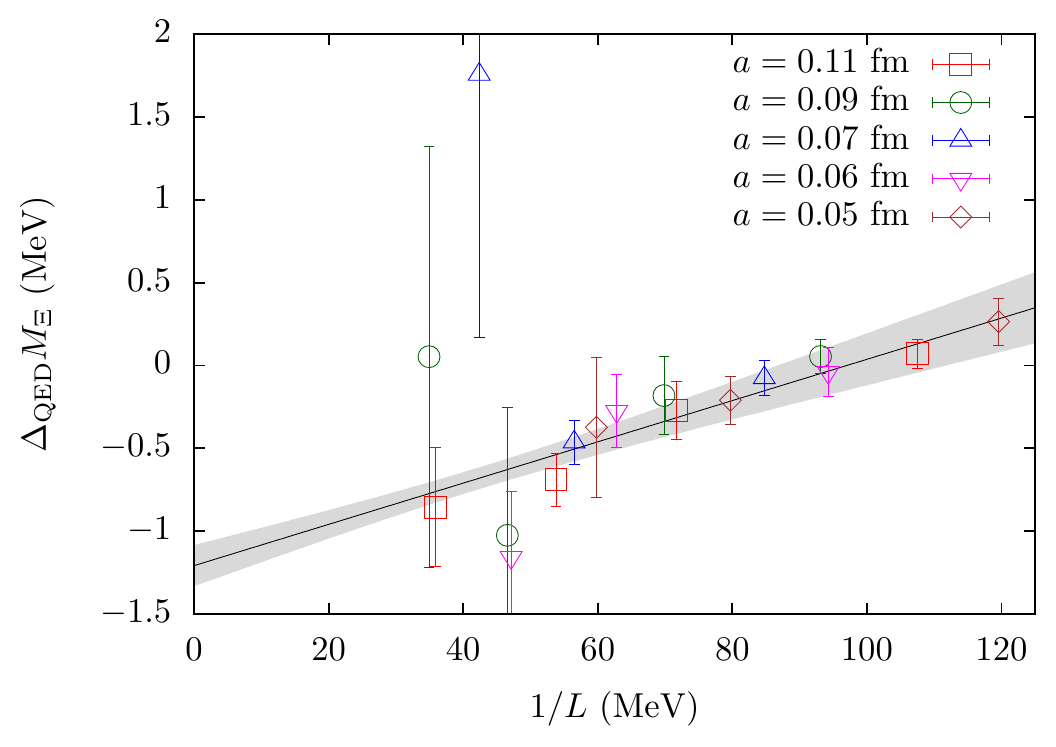}%
\caption{\label{fig:DMXiFV}\sl 
  Example of FV corrections to $\Delta_\qed M_\Xi$, plotted as a function of
  $1/L$. The dependence of the lattice results on all other variables has been
  subtracted using a fit of the type described in the text. Results with a same
  $1/L$ and $a$ are averaged because they show no systematic residual
  dependence on the other simulation parameters, in particular on quark mass.
  The linear fit in $1/L$, which is performed for points with $M_{\pi^+}\le
  500~\mev$, has a $\chi^2/\mathrm{DOF}=59./67$. 
  It is plotted as a solid curve, with its 1$\sigma$ band.}
\end{figure}

We must also account for discretization and finite-volume (FV) effects. The
latter are particularly important because of the presence of the massless
photon. Using techniques from \cite{Hayakawa:2008ci}, and performing
appropriate asymptotic expansions, it is straightforward to show that the
leading finite-volume term in scalar and spinor QED is proportional to $1/L$.
We find these corrections to be generically large. For instance, in boxes with
$L=1.6\text{--}2.6\,\fm$, which is the largest range of sizes considered in all
but the preliminary work of \cite{Basak:2013aa} on pseudoscalar masses, the
correction to $\Delta_\qed M_\Xi$, the QED contribution to $\Delta M_\Xi\equiv
M_{\Xi^0}-M_{\Xi^-}$, ranges from 123\% to 76\%. This is illustrated in
\fig{fig:DMXiFV}. In our calculation, $L$ extends up to $6~\fm$, where the
figure indicates a 36\% FV correction. While still large, our corrections are
sufficiently small that they may be described with a low-order polynomial in
$1/L$. This is confirmed by the data in \fig{fig:DMXiFV}, which show no
sensitivity to terms beyond linear order in $1/L$. The same is true of our
results for $\Delta M_N\equiv M_p-M_n$, which have a slope in $1/L$ which is
compatible with that of $\Delta_\qed M_\Xi$, but with larger statistical
errors. Not surprisingly, the slope in $\Delta M_\Sigma\equiv \Delta_{[\Delta
I_3=2]} M_{\Sigma}=M_{\Sigma^+}-M_{\Sigma^-}$ is consistent with zero: the
absolute values of the two particles' charges are equal.

Concerning discretization effects, the improvement of the QCD action implies
$O(\alpha_sa,a^2)$ corrections to $A_X$ and $B_X$. However, due to the lack of
improvement in the QED sector, discretization effects on $A_X$ are $O(a)$. In
our analysis, we include $O(a)$ QED discretization effects to $A_X$ as well as
$O(\alpha_s a,a^2)$ QCD ones to $B_X$.

Combining all of this information yields a nine parameter description of each of
the mass splittings. In the notation of \eq{eq:DMXisoexp}, this corresponds to
\bea
\label{eq:AXexp}
A_X&=&a_0^X+a_1^X[M_\pi^2-(M_\pi^{\phys})^2]+a_2^X[M_{K^\chi}^2
-(M_{K^\chi}^{\phys})^2]\nonumber\\
&&+a_3^Xa+a_4^X\frac1L\ ,\\
\label{eq:BXexp}
B_X&=&b_0^X+b_1^X[M_\pi^2-(M_\pi^{\phys})^2]+b_2^X[M_{K^\chi}^2
-(M_{K^\chi}^{\phys})^2]\nonumber\\
&&+b_3^Xf(a)\, 
\eea
where the $a_i^X$ and $b_i^X$ are the parameters and $f(a)=\alpha_s a$ or
$a^2$, alternatively. For each splitting, among the nine possible parameters, we
have retained all combinations which are such that adding one more dependence
to the fit causes the associated parameter to be consistent with zero within
one standard deviation.

\medskip
\noindent {\em Error estimation--} Our analysis methodology makes no assumptions
beyond those of the fundamental theory, except for the isospin symmetry which
is maintained in the sea and whose consequences we discuss below. However the
analysis does depend on several choices that can be sources of systematic
uncertainties.

To deal with these uncertainties, we proceed with the method put forward in
\cite{Durr:2008zz}. More specifically, we consider the following variations in
our analysis procedure. For the time ranges of the correlator fits, we consider
two initial red fit times, one for which we expect negligible excited state
contributions and a red second, more aggressive one. This estimates the
uncertainty due to contributions from excited states. Regarding the choice of
scale setting quantities, we consider two possibilities: the mass of the
$\Omega^-$ and that of the isospin averaged $\Xi$. To estimate the uncertainty
associated with the truncation of the Taylor expansion used to interpolate
these two masses to physical $M_{\pi^+}$, we vary the fit ranges by excluding
all data with pion mass above 400 and 450~MeV. To estimate part of this same
uncertainty for the isospin splittings, we consider cuts at $M_{\pi^+}=450$ and
500~MeV, since their $M_{\pi^+}^2$ dependence is very mild. These cuts also
provide an estimate of the uncertainty associated with FV corrections, as our
simulations keep $LM_\pi\sim 4$, implying cuts on $1/L$ as low as
$1/L<100~\mev$. Part of the uncertainty associated with the continuum
extrapolation is determined by considering either $\alpha_s a$ or $a^2$
discretization errors. Finally, to estimate any additional uncertainty arising
from the truncation of these expansions, we consider the result of replacing
either $A_X$ or $B_X$ by Pad\'e expressions. These are obtained by considering
that the expansions of $A_X$ and $B_X$ in Eqs.\ (\ref{eq:AXexp}) and
(\ref{eq:BXexp}) are the first two terms of a geometric series which we resum.
This resummation is not applied to the FV corrections. Instead, we try adding a
$1/L^2$ term to either the Taylor or Pad\'e forms. In all cases, we find the
coefficient of this term to be consistent with zero.

These variations lead to $2^7=128$ different fits for each of the isospin
splittings and parameter combinations. Correlating these with the 128 fits used
to determine $\Delta M^{2,\phys}$ and allowing various parameter combinations
but discarding fits with irrelevant parameters, we obtain between 64 and 256
results for each observable. The central value of a splitting is then the mean
of these results, weighted by the $p$-value. The systematic error is the
standard deviation. Because we account for all correlations, these fit
qualities are meaningful. The whole procedure is repeated for 2000 bootstrap
samples and the statistical error is the standard deviation of the weighted
mean over these samples. We have also checked that the results are changed only
negligibly (far less than the calculated errors) if they are weighted by 1
instead of by the $p$-value.

The $\delta m$ corrections that we do not include in the sea are NLO in isospin
breaking and can safely be neglected. The neglected $O(\alpha)$ sea-quark
contributions break flavor $\SU(3)$. Moreover, large-$N_c$ counting indicates
that they are $O(1/N_c)$. Combining the two suppression factors yields an
estimate $(M_\Sigma-M_N)/(N_c M_N)\simeq 0.09$. A smaller estimate is obtained
by supposing that these corrections are typical quenching effects
\cite{Aoki:2002fd} that are $\SU(3)$ suppressed, or by using
\cite{Portelli:2012pn} the NLO $\chi$PT results of \cite{Bijnens:2006mk}.
However, in the absence of direct quantitative evidence, it is safer to assume
that the EM contributions to the splittings carry an $O(10\%)$ QED quenching
uncertainty.

\begin{table}
\begin{tabular}{lccc}
\hline \hline $X$ & $\Delta M_X$ & $\Delta_\qed M_X$ & $\Delta_\qcd
M_X$\\ \hline $N$ & $\errlist{-0.68}{39}{36}$ &
$\errlist{1.59}{30}{35}$ &
$\errlist{-2.28}{25}{7}$\\ $\Sigma$ &
$\errlist{-7.84}{87}{72}$ &
$\errlist{0.08}{12}{34}$ &
$\errlist{-7.67}{79}{105}$ \\ $\Xi$ &
$\errlist{-7.16}{76}{47}$ & $\errlist{-1.29}{15}{8}$ &
$\errlist{-5.87}{76}{43}$ \\ \hline \hline
\end{tabular}
\caption{\label{tab:final_split}\sl
  Isospin breaking mass differences in MeV for members of the baryon octet. The
  first error is statistical and the second is systematic. As discussed in the
  text, we guesstimate the QED quenching uncertainties on the EM
  contributions to be $O(10\%)$. Propagating the uncertainty in $\Delta_\qed
  M_K^2$ yields an $O(4\%)$ error on the $\delta m$ contributions. The
  quenching uncertainties on the total splittings can then be obtained by
  adding those of the EM and $\delta m$ contributions in quadrature. These
  guesstimates are not included in the results.}
\end{table}

\begin{figure}[t]
\includegraphics[width=\columnwidth]{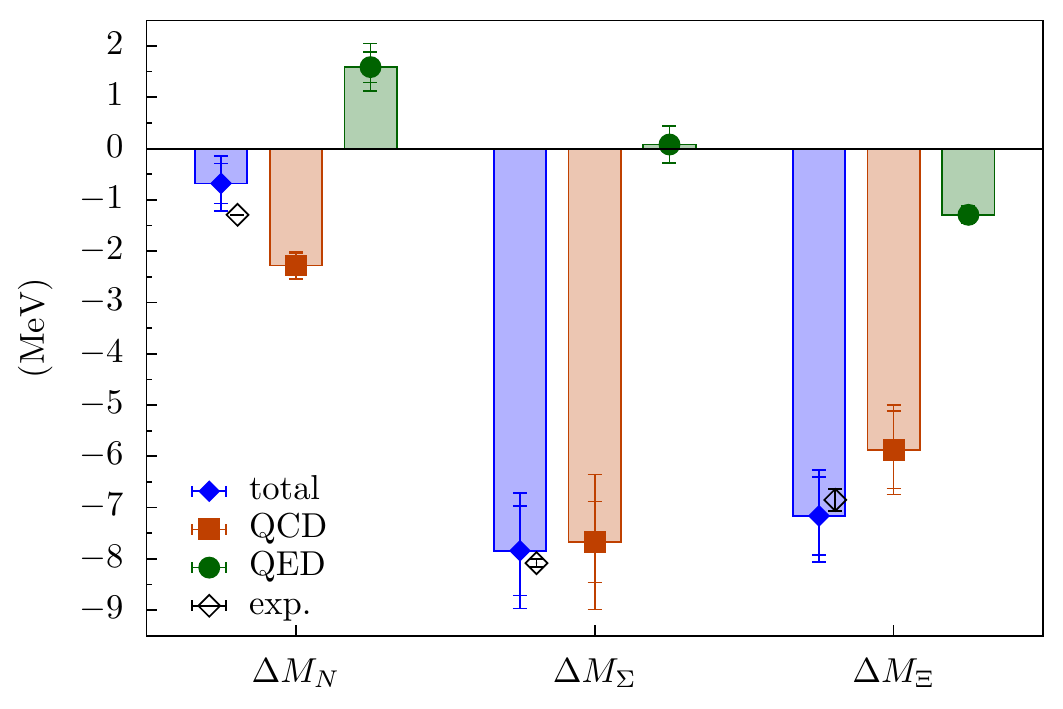}%
\caption{\label{fig:final_split}\sl
  Results for the isospin mass splittings of the octet baryons (total), the
  individual contributions to these splittings from the mass difference
  $m_u-m_d$ (QCD) and from EM (QED). The bands denote the size of these
  results. The error bars are the statistical and total uncertainties
  (statistical and systematic combined in quadrature). For comparison, the
  experimental values for the total splittings are also displayed.}
\end{figure}

\medskip
\noindent {\em Final results and discussion.--} Combining the methods described
above, we obtain our final results for the total octet baryon isospin
splittings $\Delta M_N$, $\Delta M_{\Sigma}$, and $\Delta M_{\Xi}$ defined
above. These results, together with those obtained for the EM and $\delta m$
contributions, are summarized in \tab{tab:final_split}. We also plot them in
\fig{fig:final_split}, with the experimental values for the full splittings.
Our results are compatible with experiment.

Concerning the separation into $\delta m$ and EM contributions, there exist
very few determinations of these quantities up to now. In the review
\cite{Gasser:1982ap}, hadron EM splittings were estimated using a variety of
models and Cottingham's formula for the nucleon. These estimates are compatible
with our results within $\sim 1.5\sigma$. The EM nucleon splitting has
recently been reevaluated with Cottingham's formula in
\cite{WalkerLoud:2012bg}, yielding a result which is in better agreement with
ours. $\Delta M_N$ has further been studied with sum rules in
\cite{Nasrallah:2012xz}.

Besides the entirely quenched, pioneering work of \cite{Duncan:1996be}, ours is
the only one in which the baryon octet isosplittings are fully computed. The
only other lattice calculation of the full nucleon splitting is presented in
\cite{Blum:2010ym}. Like ours, it implements QED only for valence quarks. While
their $\Delta_{\qcd}M_N$ agrees very well with ours, agreement is less good for
the EM contribution and total splitting, which they find to be
$0.38(7)~\mev$ and $-2.1(7)~\mev$, respectively. That study was performed in
rather small volumes with a limited set of simulation parameters, making an
estimate of systematic errors difficult. The few other lattice calculations
consider only the $\delta m$ contributions to the baryon splittings, in
$N_f{=}2$ \cite{deDivitiis:2011eh,deDivitiis:2013xla} and $N_f{=}2{+}1$
\cite{Beane:2006fk,Horsley:2012fw} simulations. The results of
\cite{deDivitiis:2011eh,Beane:2006fk,Horsley:2012fw} rely on imprecise
phenomenological input to fix $m_u/m_d$ or $(m_u-m_d)$. The estimate for
$\Delta_{\qed} M_K^2$ of \cite{Colangelo:2010et} is used directly in
\cite{deDivitiis:2011eh,Horsley:2012fw} and that of \cite{Aubin:2004fs},
indirectly in \cite{Beane:2006fk}. The most recent $N_f{=}2$ calculation
\cite{deDivitiis:2013xla} actually determines $\Delta_{\qed} M_K^2$ in quenched
QED, as we do here for $N_f{=}2{+}1$. $\Delta_{\qcd} M_N$ is computed in
\cite{Beane:2006fk,deDivitiis:2011eh,deDivitiis:2013xla} while all three QCD
splittings are obtained in \cite{Horsley:2012fw}. The latter is also true in
\cite{Shanahan:2012wa}, where $N_f{=}2{+}1$ lattice results are combined with
$\SU(3)$ $\chi$PT and phenomenology. Agreement with our results is typically
good. In all of these calculations, the range of parameters explored is smaller
than in ours, making it more difficult to control the physical limit.

\medskip

\begin{acknowledgments}
L.L. thanks Heiri Leutwyler for enlightening discussions. We also thank Jérôme
Charles and Marc Knecht for helpful conversations. Computations were performed
using the PRACE Research Infrastructure Resource JUGENE at FZ J\"ulich, with
further HPC resources provided by GENCI-[IDRIS/CCRT] (Grant No.\ 52275) and FZ
J\"ulich, as well as using clusters at Wuppertal and CPT. This work was
supported in part by the OCEVU Labex (ANR-11-LABX-0060) and the A*MIDEX Project
(ANR-11-IDEX-0001-02) funded by the ``Investissements d'Avenir'' French
government program managed by the ANR, by CNRS Grants GDR No.\ 2921 and
PICS No.\ 4707, by EU Grants FP7/2007-2013/ERC No.\ 208740 and No.\
MRTN-CT-2006-035482 (FLAVIAnet), and by DFG Grants No.\ FO 502/2 and No.\
SFB-TR 55.
\end{acknowledgments}

\bibliography{barsplit}

\end{document}